# Transport-induced inversion of screening ionic charges in nano-channels

Xin Zhu, Lingzi Guo, Sheng Ni, Xingye Zhang, Yang Liu*

College of Information Science and Electronic Engineering
Zhejiang University
Hangzhou, China 310013
*E-mail: yliu137@zju.edu.cn

**Abstract:**

This work reveals a counter-intuitive but basic process of ionic screening in nano-fluidic channels. Steady-state numerical simulations and mathematical analysis show that, under significant longitudinal ionic transport, the screening ionic charges can be locally inverted in the channels: their charge sign becomes the same as that of the channel surface charges. The process is identified to originate from the coupling of ionic electro-diffusion transport and junction 2-D electrostatics. This finding may expand our understanding on ionic screening and electrical double layers in nano-channels. Furthermore, the charge inversion process results in a body-force torque on channel fluids, which is a possible mechanism for vortex generation in the channels and their nonlinear current-voltage characteristics.

**TOC Graphic**

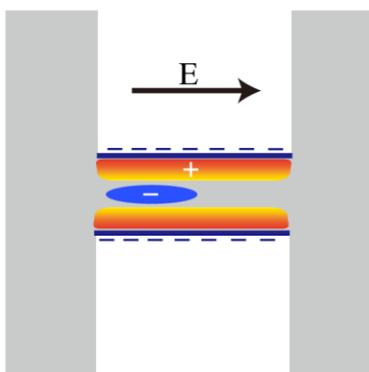

Ionic screening is a fundamental physical process in electrolytic solutions. In nano-fluidic channels, the formation of screening electrical double layers (EDLs) at channel surfaces plays a central role in their transport phenomena, including perm-selectivity[1-3], concentration polarization[4-8], and nonlinear conductance[9-20], as well as their broad device applications[21-26]. In modeling ionic screening in nano-channels, the condition of local electroneutrality was often adopted by assuming complete screening in the channel transversal direction: the sum of the surface and ion charges along an arbitrary transversal line is zero[7,8,13]. Nevertheless, more detailed account of the electrostatics, particularly at the junctions of nano-channels and micro-reservoirs, have been found important when electrokinetic transport is significant [10,14,15,21]. The existence of the junction space charge regions and their extending into the reservoirs were revealed and proposed to explain the over-limiting ionic currents [10,11,14]. At such junctions, counter-ions diffuse into the reservoirs and leave behind under-screened channels where the surface charges are not fully screened. They are analogous to semiconductor junctions, and nano-fluidic junction devices have been proposed accordingly[21,22]. The impact of transport on nano-channel's ion selectivity has been revealed[3]. Recently, a semi-analytical capillary pore model based on local quasi-equilibrium has been developed for electrolyte transport through nanopores[27]. For side-gated nanopores, the transport-induced suppression of ionic screening, i.e., the descreening effect, has also been numerically investigated[28,29]. Thus far, a commonly-accepted basic concept on ionic screening is that the screening ionic charges should have an opposite sign to that of the surface charges, since the former are induced to counteract the effect of the latter.

This paper reports that, under significant ionic transport, the sign of the screening ionic charges can be locally inverted in the nano-channels and become the same as that of the surface charges. The study is based on numerical simulations and mathematical analysis of the continuum transport model (Poisson-Nernst-Planck and Stokes). Although counter-intuitive, the charge inversion process is found not a peculiarity, but a rather general effect resulting from the coupling of the ionic electro-diffusion transport and junction electrostatics. The finding may expand our understanding on ionic screening in nano-channels as well as imply device applications. Particularly, it indicates that the ion selectivity of nano-channels is no longer solely defined by the channel surface charges and can even be locally inverted by longitudinal ionic transport. Furthermore, the impact of the charge inversion process on the fluid transport, particularly its relation to channel vortex generation and nonlinear conductance, is investigated.

We firstly model a typical nano-channel as schematically shown in Fig. 1(a). The channel is a nano-slit geometry with a negative surface charge density $-\sigma_s$. The channel connects two micro-reservoirs filled with a KCl solution of bulk concentration $c_0$. An electrical bias, $V_d$, is applied between the two reservoirs and generates an ion current, $I_d$. Self-consistent numerical simulations are conducted based on the steady-state Poisson-Nernst-Planck-Stokes (PNP-S) equations[28,29] (see Supporting Information). The fluid transport is modeled as a low-Reynolds-number flow with hydrophilic, no-slip boundary condition[30]. The COMSOL MultiPhysics software is used for simulations and calibrated with previously reported results. The bulk ion concentration $c_0$ is 1mM, corresponding to a Debye length $\Lambda_D$ of 10nm. The default channel height $h$ is 100nm~$10\Lambda_D$. Other default simulation parameters include $l = 1\mu m$, $H' = 2.1\mu m$, $L' = 1\mu m$. $-\sigma_s$ is set to $-0.01 q/nm^2$. $\mu$ is $7.62 \times 10^{-8} m^2/Vs$ for either ion species. For the ion diffusivity $D$, the Einstein relation is used $D = \mu k_B T/q$, where $k_B$ is the Boltzmann constant, and $T$ the temperature. The viscosity $\gamma$ is $10^{-3} Ns/m^2$.

The phenomenon of screening charge inversion is clearly shown in Fig. 2a, where the normalized density of net ionic charges, $(c_+ - c_-)/c_0$, is plotted for a $V_d$ of 12V. Here, $c_+$ and $c_-$ are the concentrations of cations and anions, respectively. It is shown that positive ionic charges are induced adjacent to the negatively charged channel surface, as commonly expected for EDLs. However, at the center of the channel, a region of negative ionic charges forms near the junction on the left side. This is in direct contrast to our common understanding that positive ionic charges should maintain at the overlapping EDL tails throughout the channel.

Taking a closer look, the transversal distributions of $c_+$ and $c_-$ are plotted in Fig. 2b along a line near the junction ($z = 250 nm$). At zero bias, they are shown to follow the Poisson-Boltzmann equation: both tend to the bulk value $c_0$ toward the channel center; $c_+$ is always greater than $c_-$ throughout the channel. At 12V bias, however, the counter-ion concentration $c_+$ tends to a lower asymptote than the co-ion concentration $c_-$ does; $c_+$ is clearly smaller than $c_-$ in the center region ($0 \leq x \leq 30 nm$).

The dependence of ionic charge inversion on channel heights and biases is further examined in Fig. 2c. Four channel heights are simulated corresponding to $6\Lambda_D$, $8\Lambda_D$, $10\Lambda_D$, and $15\Lambda_D$, respectively, while the surface charge density is kept constant. For each case, the normalized net ionic charges are plotted along the centerline (x=0) under various biases. As $V_d$ increases, the distribution becomes qualitatively different from the zero-bias, Poisson-Boltzmann solution. The initially positive net ion charges are gradually reduced and become inverted to negative in a significant portion of the channel. It is further observed that the charge inversion is the most evident for moderately coupled EDLs ($h \sim 8\Lambda_D$). As the channel height becomes smaller (i.e., stronger EDL overlap), higher biases are needed to reach the inversion condition. On the other hand, as the channel height becomes greater (i.e., weaker EDL overlap), the inversion condition is reached at lower $V_d$, but the magnitude of the inverted charges becomes less significant. Such dependence will be explained subsequently in this paper. Further simulations are conducted for different values of parameters including $\sigma_s$ and $c_0$, and the charge inversion phenomenon occurs in general (see Supporting Information). Additionally, cylindrically symmetric nanopores with various pore diameters are simulated and presented in the Supporting Information. The charge inversion phenomenon is evident in the cylindrical channels as well, but its scaling with channel dimension is quantitatively different from that of the nano-slits due to their different geometries.

The ionic charge inversion can directly impact the transport properties of nano-channels. In particular, it produces a possible mechanism for vortex generation inside the nano-channels. It is known that the product of the net ion charges and electric field determines the electrical body force on fluids. When driven by significant biases, the longitudinal electric field points to the same direction throughout the channel. Consequently, the charge inversion region exerts a body force opposite to that at the surface, thus applying a torque to the channel fluid. The body-force torque increases with $V_d$ and can result in channel vortices at high biases. This effect is shown in Fig. 3(a)(b), where simulated fluid flow patterns are plotted for moderate (5V) and high (15V) biases. At 5V, the fluid flow exhibits a typical electro-osmotic flow (EOF) that is parallel to the channel surface. In contrast, a recirculating vortex is generated inside the channel at 15V due to the body-force torque.

The fluidic flow patterns in turn have significant impact on the channel conductance and may lead to strong current-voltage nonlinearity. As shown in Fig. 3(c), the simulated current-voltage curve clearly exhibits a suppression stage at medium biases and a restoration stage at high biases. The two stages are associated with the two flow patterns in Fig. 3(a) and (b), respectively. Similar observations have also been found for electrically gated nanopores[28,29]. Our recent theoretical study has identified that the channel vortices are the underlying cause of the conductance restoration, but the mechanism of vortex generation has not been identified[31]. The ionic charge inversion revealed in this work offers an explanation to the origin of the channel vortex and its resultant conductance restoration in these studies.

An important question to answer is whether charge inversion is essentially caused by the fluid transport, particularly the vortices, rather than the cause of the latter. This necessity is ruled out based on our additional simulations using the Poisson-Nernst-Planck (PNP) model: in such a case, the Stokes equation is not included, i.e. the fluid transport is not modeled, but the screening charge inversion is still observed (see Supporting Information). It means that the root cause of charge inversion is contained in the PNP model. Note that this does not exclude the possibility that fluid flow can affect the generated charge inversion. In fact, electro-osmosis has been shown to alter the ion distributions in nano-channels with weakly coupled EDLs[31]; its effect on ionic charge distribution is expected. Nonetheless, the following analysis intends to identify the root cause of charge inversion and therefore will be based on the PNP model.

To reveal the root cause, a simplified structure (Fig. 1b) is studied using the PNP model. This simplified case captures the essential physics for charge inversion while facilitates our mathematical analysis. In the following, we show that two factors are key to the creation of inverted screening charges at high biases: the extending of the under-screening regions and the dominance of longitudinal electrostatics over the transversal one.

The first key factor is illustrated in Fig. 4a, where the bias dependence of $\overline{E_z}$ is plotted. Here, $\overline{E_z}$ is an average quantity defined as $2/h \int_0^{h/2} E_z(x,z)dx$, where $E_z(x,z)$ is the simulated 2-D profile of longitudinal electric field. Under zero bias, $\overline{E_z}$ at the two junctions shows the typical exponential decay that is characterized by the Debye length. As $V_d$ increases, the profiles of $\overline{E_z}$ are qualitatively different, showing significantly extended junctions with non-trivial gradient of $\overline{E_z}$. According to Gauss's law, this corresponds to net space charges extending to both the channel and the reservoirs.

To further analyze the extending effect, we can average the PNP equations along the transversal direction and obtain the following 1-D homogenized equations:

$$-D\, d\overline{c_\pm}/dz \pm \mu \overline{c_\pm} \cdot \overline{E_z} = \overline{f_\pm}, \qquad (1)$$

$$\varepsilon_w d\overline{E_z}/dz = \bar{\rho}, \qquad (2)$$

where $\varepsilon_w$ is the solution permittivity, $\mu$ mobility, $D$ diffusivity, and $f_\pm$ cation and anion flux densities, respectively. The overbar denotes that the corresponding quantity is averaged over the transversal direction. The ion fluxes, $\overline{f_\pm}$, are constant due to continuity. In deriving these 1-D equations, the assumption, $\overline{c_\pm \cdot E_z} \approx \overline{c_\pm} \cdot \overline{E_z}$, has been made, which is valid for weak EDL overlap (i.e. $\Lambda_D \ll h$)[13] or under large biases. The average net charge density $\bar{\rho}$ includes contributions from both ion and surface charges. It is $q(\overline{c_+} - \overline{c_-}) - 2\sigma_s/h$ inside the channel ($0 \leq z \leq l$) and $q(\overline{c_+} - \overline{c_-})$ in the reservoirs. Accordingly, the entire structure consists of three segments, intrinsic reservoir/positive channel/intrinsic reservoir, as illustrated in Fig. 1c. Within each segment, the following equation for $\overline{E_z}$ can be readily derived:

$$-\frac{d^2 \overline{E_z}}{dz^2} + \frac{q^2}{k_B T \varepsilon_w}(\overline{c_+} + \overline{c_-}) \cdot \overline{E_z} = \bar{J}/D\varepsilon_w, \qquad (3)$$

where $\bar{J} = q(\overline{f_+} - \overline{f_-})$ is the net current density. The two terms on the left hand side correspond to the diffusion and drift current components, respectively.

Under equilibrium, the net current is zero. The solution to Eq. 3 therefore leads to the exponential dependence typical for Debye screening, $\overline{E_z} \sim exp\left(\pm(z - z_0)/\Lambda_D\right)$, at $z_0 = 0 \: or \: l$. However, for non-zero biases, the solution to Eq. 3 in general consists of more slowly varying spatial dependence. For example, in the region where the drift component dominates over the diffusion one, $\overline{E_z}$ varies as $\sim 1/\overline{c_+} + \overline{c_-}$. Consequently, $\bar{\rho}$ varies as $\sim \frac{d}{dz}\left(1/\overline{c_+} + \overline{c_-}\right)$, which is no longer characterized by $\Lambda_D$ [31].

For small $\sigma_s$, an analytical expression for $\overline{E_z}$ can be derived from a first-order perturbation analysis on the PNP equations (see Supporting Information). The solution consists of a superposition of the Debye exponential decay and a linear component. When the approximation, $\left(q\Lambda_D V_d/k_B T L_t\right)^2 \ll 1$, is satisfied, the solution can be simplified as given in the Methods. In Fig. 4b, the analytical expression is compared against the simulation results; a good match is observed for small $\sigma_s$. The linear component in $\overline{E_z}$ corresponds to constant net space charges, $-2\varepsilon_w kL'$ and $\varepsilon_w kl$, in the entire channel and reservoirs, respectively.

Here, $k \approx V_d^2 \sigma_s / c_0 k_B T L_t^3 h$ and $L_t \equiv l + 2L'$. As $\sigma_s$ increases, the analytical expression starts to deviate from simulation results due to higher-order effects. In particular, $\bar{\rho}$ is no longer constant in the channel. But it still remains negative in an extended portion of the channel.

In the next, we examine the essential role of the second key factor, i.e. the dominance of the longitudinal electrostatics. The following analysis is general and not limited to the perturbation analysis. It is important to understand that the first key factor, i.e. the extended under-screening effect, alone is not sufficient for creating the local charge inversion. Thus far, our 1-D homogenized model above only shows that the averaged net charge density, $\bar{\rho} \equiv q(\overline{c_+} - \overline{c_-}) - 2\sigma_s/h$, is negative in the channel. After excluding the contribution of the negative surface charges, the averaged ionic charge density, $\bar{\rho}_i \equiv q(\overline{c_+} - \overline{c_-})$, is still positive throughout the channel and not inverted. This does not contradict our observation in Fig. 2, where the ionic charge inversion only occurs locally around the center. Therefore, what our analysis needs to show is that the local ionic charge density along the centerline, $\rho_{i,c} \equiv q[c_+(x=0,z) - c_-(x=0,z)]$, can become negative.

The 2-D Poisson's equation along the centerline is

$$\varepsilon_w \left[ dE_z(x=0,z)/dz + dE_x(x=0,z)/dx \right] = \rho_{i,c}. \quad (4)$$

Here, the two left-hand-side terms, $dE_z/dz$ and $dE_x/dx$, represent the electrostatic control from the longitudinal and transversal directions, respectively. Under sufficiently high $V_d$, the magnitude of the former becomes dominant over the latter at the channel center, and the latter can therefore be omitted from Eq. 4. At the same time, the local electric field, $E_z(x=0,z)$, approaches the transversally averaged quantity, $\overline{E_z}$ (see Supporting Information). We then compare Eq. 4 with the average 1-D Poisson's equation (Eq. 2) and can readily find that, because their left-hand-side terms approach each other, $\rho_{i,c}$ should approach $\bar{\rho}$ as well. Since $\bar{\rho}$ is already shown above to be negative due to the extended under-screening, it therefore follows that $\rho_{i,c}$ also tends to be negative in the channel, i.e. the ionic charge inversion occurs locally along the centerline.

It is worth emphasizing that, in the above analysis, what $\rho_{i,c}$ approaches is $\bar{\rho}$, not $\bar{\rho}_i$. In the process of under-screening, $\bar{\rho}_i$ is positive but smaller than $2\sigma_s/h$, and $\bar{\rho}$ is negative. To verify the analysis, the local and average net charges, $\rho_{i,c}$ and $\bar{\rho}$, are compared in Fig. 5. As $V_d$ increases, it is evident that the former approaches the latter more closely. The physics of local ionic charge inversion can therefore be summarized as the combined effect of the two key factors: the ionic transport makes surface charges under-screened; the average net charges is therefore negative; the average longitudinal field distributes according to the negative net charges; the local longitudinal field follows the average one; when the transversal field effect is negligible, this local longitudinal field induces negative local ionic charges following Gauss's law.

Following the above reasoning, we have $|\bar{\rho}| \equiv \left|\bar{\rho}_i - 2\sigma_s/h\right| \leq \left|2\sigma_s/h\right|$; the maximum magnitude is reached when the surface charges are completely un-screened. Since $\rho_{i,c}$ follows $\bar{\rho}$, its magnitude does not exceed $\left|2\sigma_s/h\right|$ as well. This observation then explains the dependence on channel heights mentioned early in this paper. For small channels with strong EDL overlap, it takes higher biases for $\left|dE_z/dz\right|$ to dominate over the elevated $\left|dE_x/dx\right|$ at the channel center. On the other hand, for very large channel heights, the magnitude of inverted charges becomes less significant because of the reduction in $\left|2\sigma_s/h\right|$ for fixed $\sigma_s$.

The remaining of this Letter further discusses several aspects related to the inversion process. Fluidic vortices have been frequently observed in nano-channel structures, particularly in the reservoirs next to the channel entrances[6,12,32]. Their origins have been intriguing, and various mechanisms have been proposed in the past, including electrokinetic instability[11-13], varying effective electric fields and induced pressure gradient[8,33], EOF around corners[15,34], and EOF backflow[13]. The body-force torque revealed in this work is intrinsically different from those mechanisms. In particular, the vortex modeled by Mani et al. is different from this work since their model was based on local electroneutrality[8]. Yossifon et al. identified another type of steady-state vortex and attributed its origin to the sharp bending of EOF flow around the channel corners[34,35]. That analysis relied on the condition that the micro-reservoir walls also have significant surface charges[34]. In our previous[31] and present studies, we have respectively modeled the cases with and without micro-reservoir surface charges. In both cases, the vortex is generated inside the channel, indicating that it is not essentially caused by EOF bending at the corners.

An experimental observation of intra-channel vortices was previously reported by visualizing the nano-colloid transport within nano-channels[19]. In that work, streamlines in opposite directions at one end of the channel can be observed, indicating the formation of a vortex pair. It was also observed that the nano-colloids pack and unpack in the channel periodically. Whether the charge inversion process is responsible for the observed intra-channel vortices in that work remains to be further examined. Since the packed nano-colloid structure may alter the channel electric field[19], the model would need to self-consistently account for the effects of the packed nano-colloids, and it would be interesting to explore how the nano-colloids can modulate the charge inversion process.

In a previous work on concentration polarization, the positive feedback between depleted ion concentration and increased local Debye length has been studied[6]. However, we note that the increased Debye length at the ion depletion side does not fully account for the extended under-screening, as evident from the perturbation analysis in Methods. In this case, the ion concentration is only slightly perturbed from $c_0$ (1mM), so the local Debye length is still very close to its bulk value (10nm). But our results clearly show that net space charges can extend to the entire channel (1$\mu m$ long), far beyond the local Debye length. This is consistent with the previously studied descreening effect, in which the field effect is shown to extend beyond the limit of Debye lengths[29,36].

The under-screening process is closely related to the extending of space charges into the reservoirs, which was previously proposed to explain the overlimiting currents and the instable generation of vortices[10,11,14]. However, the space charges studied in those works were located in the reservoirs and positive for cation-selective channels. Therefore, they were still based on the common ionic screening picture. In another closely related study, the nano-channel's ion selectivity was found to change with ionic transport, but the applied bias was not high enough to reach charge inversion due to strong EDL overlap of the channels[3]. Overall, the charge inversion effect in this paper has not been revealed in those related works.

More insight into the cause of the charge inversion process may be obtained by making a comparison with semiconductor junctions as schematically shown in Fig. 1c. As a positive $V_d$ is applied, the left i-p junction is reversely biased, and the right p-i junction forwardly biased. The left junction region is therefore widened, analogous to the extended under-screening process. However, there are important differences to be noted. In semiconductor junctions, the electrostatic potential is parabolic in the fully depletion zone and becomes the Debye type at the zone edges[37]. In the ionic solutions, if one only considers the simplified 1-D model, the homogenized surface charges play the role of dopants, and the PNP equations are almost identical to the semiconductor Shockley equations. An exception is that the former does not have the recombination term for strong electrolytes[22]. Such a lack of the law of mass action indicates infinitely long diffusion lengths in the PNP model. Therefore, despite the separation by the long channel in between, the i-p junction and the p-i junction are still coupled electrostatically, as observed in Fig. 4. This is in contrast to the case of two back-to-back semiconductor diodes that are separated further than their finite carrier diffusion lengths. For this reason, the under-screening regions extend more significantly in the case of nano-channels.

The most fundamental difference from the semiconductor junctions comes from the fact that the effective "doping" from channel surface charges is not truly volumetric doping. For truly volumetric doping such as semiconductors or ionic solutions with polyelectrolyte gels[38], the screening charge inversion does not occur: the dopant charges co-exist with screening charges at each location; the under-screening is achieved when the magnitude of the latter is smaller than that of the former. On the other hand, for nano-channels effectively "doped" by surface charges, there are no actual "dopant" charges away from the surfaces; the local ionic charges have to be inverted to satisfy the under-screening electrostatics.

Another type of charge inversion phenomenon has been extensively studied for electrolytes with multivalent ions[39]. Its underlying reason is the over-screening of the surface charges by counter-ions, which makes the over-all net charges opposite to the surface charges. It is of a completely different nature from the screening charge inversion process in this study.

The screening process is central to nano-channel functions, and therefore the inversion effect may have important implications on device applications. As already shown above, its effect on fluidic flow pattern can directly impact the nonlinearity of ionic conductance, a key property of nano-channels and nano-pores. Furthermore, it leads to a decreased ion perm-selectivity at high biases. For applications involving translocation of biomolecules, a local region of inverted screening charges can change the cross-section to capture the biomolecules. Moreover, the generated channel vortices may be used for mixing of solutions or local trapping of molecules.

In summary, numerical simulations and mathematical analysis based on the Poisson-Nernst-Planck-Stokes model has revealed that the screening ionic charges can be locally inverted in the channels by electro-diffusion transport of ions. Two factors have been identified as the underlying reasons of this effect: the transport-induced under-screening process and the dominance of the longitudinal electrostatics over the transversal electrostatics under high biases. The charge inversion process has also been shown to apply a body-force torque on the fluid, thus producing a mechanism for vortex generation inside the channels.

**Computational Methods:**

**Perturbation Analysis:** Starting with the 1-D PNP equations and fixing $V_d$, we can conduct a first-order perturbation analysis with respect to small $\sigma_s$. The resultant ordinary differential equations can be analytically solved, as described in the Supporting Information. The expression for $\overline{E_z}$ at small $\sigma_s$ is a superposition of the Debye exponential decay and a linear component. When the condition $\left(q\Lambda_D V_d / k_B T L_t\right)^2 \ll 1$ is satisfied, the general solution can be simplified as an approximate expression:

$$\overline{E_z} \approx \frac{V_d}{L_t} + \begin{cases} \frac{\psi_D}{2\Lambda_D} \exp\left(z/\Lambda_D\right) + k \cdot l \cdot (z + L'), & \text{if } L' \leq z < 0 \\ \frac{\psi_D}{2\Lambda_D} \exp\left(-z/\Lambda_D\right) - k \cdot 2L' \cdot \left(z - l/2\right) - \frac{\psi_D}{2\Lambda_D} \exp\left(z - l/\Lambda_D\right), & \text{if } 0 \leq z < l, \\ -\frac{\psi_D}{2\Lambda_D} \exp\left(z - l/\Lambda_D\right) + k \cdot l \cdot (z - l - L'), & \text{if } l \leq z \leq l + L' \end{cases}$$

Eq. (5)

where $L_t \equiv l + 2L'$, the Donnan potential $\psi_D \equiv k_B T/q \sinh^{-1}\left(\sigma_s/qhc_0\right)$, and $k \approx V_d^2 \sigma_s / c_0 k_B T L_t^3 h$. According to Gauss's law, the linear component in $\overline{E_z}$ corresponds to constant net space charges, $-2\varepsilon_w k L'$ and $\varepsilon_w k l$, throughout the entire channel and reservoirs, respectively. The cases discussed in Fig. 4 satisfy the condition $\left(q\Lambda_D V_d / k_B T L_t\right)^2 \ll 1$, and therefore we use this simplified equation for the calculations. More derivation details are included in the Supporting Information.

**Acknowledgements:**

Discussions with Prof. Robert Dutton at Stanford University are gratefully acknowledged. This work was supported by National Natural Science Foundation of China (Grant No: 61574126) and Natural Science Foundation of Zhejiang Province, China (Grant No: LR15F040001).

**Supporting Information:** Numerical simulation model description; additional simulation results for various parameter values, cylindrical nano-pore geometry, and the PNP model, respectively; Derivation details of perturbation analysis; comparison of average and local electric fields.

**References:**
(1) Schoch, R. B.; Han, J.; Renaud, P. Transport Phenomena in Nanofluidics. *Rev. Mod. Phys.* **2008**, *80*, 839–883.
(2) Stein, D.; Kruithof, M.; Dekker, C. Surface-Charge-Governed Ion Transport in Nanofluidic Channels. *Phys. Rev. Lett.* **2004**, *93*, 035901.
(3) Vlassiouk, I.; Smirnov, S.; Siwy, Z. Ionic Selectivity of Single Nanochannels. *Nano Lett.* **2008**, *8*, 1978–1985.
(4) Pu, Q.; Yun, J.; Temkin, H.; Liu, S. Ion-Enrichment and Ion-Depletion Effect of Nanochannel Structures. *Nano Lett.* **2004**, *4*, 1099–1103.
(5) Höltzel, A.; Tallarek, U. Ionic Conductance of Nanopores in Microscale Analysis Systems: Where Microfluidics Meets Nanofluidics. *J. Sep. Sci.* **2007**, *30*, 1398–1419.
(6) Kim, S. J.; Wang, Y.-C.; Lee, J. H.; Jang, H.; Han, J. Concentration Polarization and Nonlinear Electrokinetic Flow Near a Nanofluidic Channel. *Phys. Rev. Lett.* **2007**, *99*, 044501.
(7) Green, Y.; Shloush, S.; Yossifon, G. Effect of Geometry on Concentration Polarization in Realistic Heterogeneous Permselective Systems. *Phys. Rev. E* **2014**, *89*, 043015.
(8) Mani, A.; Zangle, T. A.; Santiago, J. G. On the Propagation of Concentration Polarization From Microchannel–Nanochannel Interfaces Part I: Analytical Model and Characteristic Analysis. *Langmuir* **2009**, *25*, 3898–3908.
(9) Levich, V. G. Physicochemical Hydrodynamics; Prentice Hall, 1962.
(10) Rubinstein, I.; Shtilman, L. Voltage Against Current Curves of Cation Exchange Membranes. *J. Chem. Soc., Faraday Trans. 2* **1979**, *75*, 231.
(11) Rubinstein, I.; Zaltzman, B. Dynamics of Extended Space Charge in Concentration Polarization. *Phys. Rev. E* **2010**, *81*, 061502.
(12) Yossifon, G.; Chang, H.-C. Selection of Nonequilibrium Overlimiting Currents: Universal Depletion Layer Formation Dynamics and Vortex Instability. *Phys. Rev. Lett.* **2008**, *101*, 254501.
(13) Dydek, E. V.; Zaltzman, B.; Rubinstein, I.; Deng, D. S.; Mani, A.; Bazant, M. Z. Overlimiting Current in a Microchannel. *Phys. Rev. Lett.* **2011**, *107*,

118301.
(14) Manzanares, J. A.; Murphy, W. D.; Mafe, S. Numerical Simulation of the Nonequilibrium Diffuse Double Layer in Ion-Exchange Membranes. *The Journal of Physical Chemistry* **1993**.
(15) Yossifon, G.; Mushenheim, P.; Chang, Y.-C.; Chang, H.-C. Eliminating the Limiting-Current Phenomenon by Geometric Field Focusing Into Nanopores and Nanoslots. *Phys. Rev. E* **2010**, *81*, 046301.
(16) Cheng, L.-J.; Guo, L. J. Rectified Ion Transport Through Concentration Gradient in Homogeneous Silica Nanochannels. *Nano Lett.* **2007**, *7*, 3165–3171.
(17) Fuest, M.; Boone, C.; Rangharajan, K. K.; Conlisk, A. T.; Prakash, S. A Three-State Nanofluidic Field Effect Switch. *Nano Lett.* **2015**, 150306111917005.
(18) Green, Y.; Eshel, R.; Park, S.; Yossifon, G. Interplay Between Nanochannel and Microchannel Resistances. *Nano Lett.* **2016**, *16*, 2744–2748.
(19) Chang, H.-C.; Yossifon, G. Understanding Electrokinetics at the Nanoscale: a Perspective. *Biomicrofluidics* **2009**, *3*, 12001.
(20) Nam, S.; Cho, I.; Heo, J.; Lim, G.; Bazant, M. Z.; Moon, D. J.; Sung, G. Y.; Kim, S. J. Experimental Verification of Overlimiting Current by Surface Conduction and Electro-Osmotic Flow in Microchannels. *Phys. Rev. Lett.* **2015**, *114*, 114501.
(21) Daiguji, H.; Oka, Y.; Shirono, K. Nanofluidic Diode and Bipolar Transistor. *Nano Lett.* **2005**, *5*, 2274–2280.
(22) Cheng, L.-J.; Guo, L. J. Nanofluidic Diodes. *Chem Soc Rev* **2010**, *39*, 923–938.
(23) Karnik, R.; Fan, R.; Yue, M.; Li, D.; Yang, P.; Majumdar, A. Electrostatic Control of Ions and Molecules in Nanofluidic Transistors. *Nano Lett.* **2005**, *5*, 943–948.
(24) Paik, K.-H.; Liu, Y.; Tabard-Cossa, V.; Waugh, M. J.; Huber, D. E.; Provine, J.; Howe, R. T.; Dutton, R. W.; Davis, R. W. Control of DNA Capture by Nanofluidic Transistors. *ACS Nano* **2012**, *6*, 6767–6775.
(25) Tagliazucchi, M.; Szleifer, I. Salt Pumping by Voltage-Gated Nanochannels. *J. Phys. Chem. Lett.* **2015**, *6*, 3534–3539.
(26) La Mantia, F.; Pasta, M.; Deshazer, H. D.; Logan, B. E.; Cui, Y. Batteries for Efficient Energy Extraction From a Water Salinity Difference. *Nano Lett.* **2011**, *11*, 1810–1813.
(27) Peters, P. B.; van Roij, R.; Bazant, M. Z.; Biesheuvel, P. M. Analysis of Electrolyte Transport Through Charged Nanopores. *Phys. Rev. E* **2016**, *93*, 053108.
(28) Liu, Y.; Huber, D. E.; Dutton, R. W. Limiting and Overlimiting Conductance in Field-Effect Gated Nanopores. *Appl. Phys. Lett.* **2010**, *96*, 253108.
(29) Liu, Y.; Huber, D. E.; Tabard-Cossa, V.; Dutton, R. W. Descreening of Field Effect in Electrically Gated Nanopores. *Appl. Phys. Lett.* **2010**, *97*, 143109.
(30) Bouzigues, C. I.; Tabeling, P.; Bocquet, L. Nanofluidics in the Debye Layer at Hydrophilic and Hydrophobic Surfaces. *Phys. Rev. Lett.* **2008**, *101*, 114503.
(31) Liu, Y.; Guo, L.; Zhu, X.; Ran, Q.; Dutton, R. Suppression of Ion Conductance by Electro-Osmotic Flow in Nano-Channels with Weakly Overlapping Electrical Double Layers. *AIP Advances* **2016**, *6*, 085022.

(32) Yossifon, G.; Chang, Y.-C.; Chang, H.-C. Rectification, Gating Voltage, and Interchannel Communication of Nanoslot Arrays Due to Asymmetric Entrance Space Charge Polarization. *Phys. Rev. Lett.* **2009**, *103*, 154502.

(33) Park, S. Y.; Russo, C. J.; Branton, D.; Stone, H. A. Eddies in a Bottleneck: an Arbitrary Debye Length Theory for Capillary Electroosmosis. *Journal of Colloid and Interface Science* **2006**, *297*, 832–839.

(34) Chang, H.-C.; Yossifon, G.; Demekhin, E. A. Nanoscale Electrokinetics and Microvortices: How Microhydrodynamics Affects Nanofluidic Ion Flux. *Annu. Rev. Fluid Mech.* **2012**, *44*, 401–426.

(35) Yossifon, G.; Chang, H.-C. Changing Nanoslot Ion Flux with a Dynamic Nanocolloid Ion-Selective Filter: Secondary Overlimiting Currents Due to Nanocolloid-Nanoslot Interaction. *Phys. Rev. E* **2010**, *81*, 066317.

(36) Liu, Y.; Sauer, J.; Dutton, R. W. Effect of Electrodiffusion Current Flow on Electrostatic Screening in Aqueous Pores. *J. Appl. Phys.* **2008**, *103*, 084701.

(37) Sze, S. M.; Ng, K. K. Physics of Semiconductor Devices; John Wiley & Sons: Hoboken, NJ, USA, 2006.

(38) Doi, M.; Yamamoto, T. Electrochemical Mechanism of Ion Current Rectification of Polyelectrolyte Gel Diodes. *Nature Communications* **2014**, *5*, 1–8.

(39) Grosberg, A. Y.; Nguyen, T. T.; Shklovskii, B. I. Colloquium: the Physics of Charge Inversion in Chemical and Biological Systems. *Rev. Mod. Phys.* **2002**.

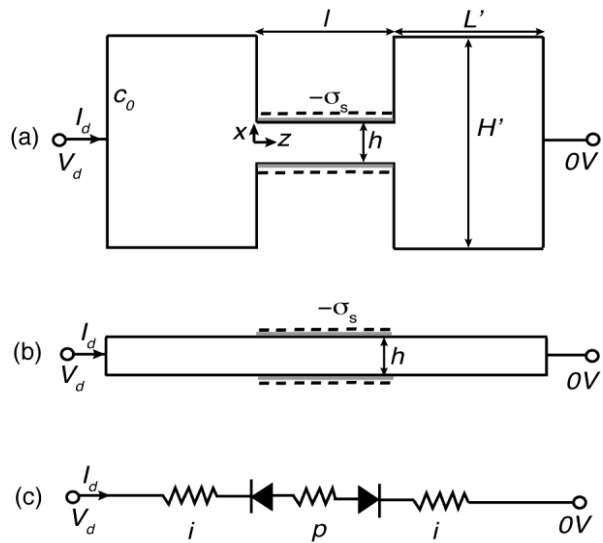

*Fig. 1: (a) The basic nano-channel structure modeled in this work. The nano-channel connects two micro-reservoirs. The negative channel surface charge density is $-\sigma_s$, the bulk KCl ion concentration $c_0$, and the applied voltage bias $V_d$; (b) Simplified nano-channel structure; (c) Semiconductor analogy illustrating the two connected junctions: reverse-biased i-p junction on the left and forward-biased p-i junction on the right.*

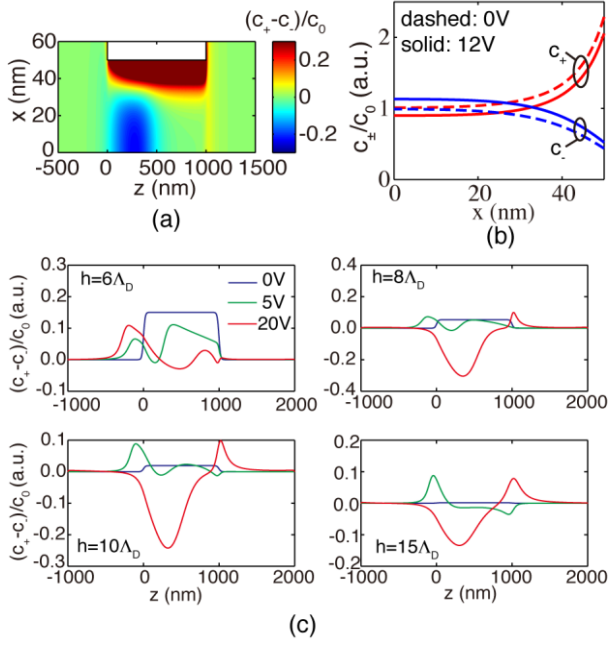

*Fig. 2 (a) Simulated 2-D distribution of normalized net ion charges, $(c_+ - c_-)/c_0$, around the channel of the structure in Fig. 1a. The model is PNP-S. $V_d$ is 12V. Only the device top half is shown; (b) Simulated transversal profiles of the normalized ion concentrations, $c_\pm/c_0$, along the line $z = 250nm$ for two biases, 0V and 12V; (c) Simulated longitudinal profiles of normalized net ion charges, $(c_+ - c_-)/c_0$, along the centerline ($x = 0$) for three biases, 0V, 5V, and 20V. Various channel heights are simulated: $6\Lambda_D$, $8\Lambda_D$, $10\Lambda_D$, and $15\Lambda_D$.*

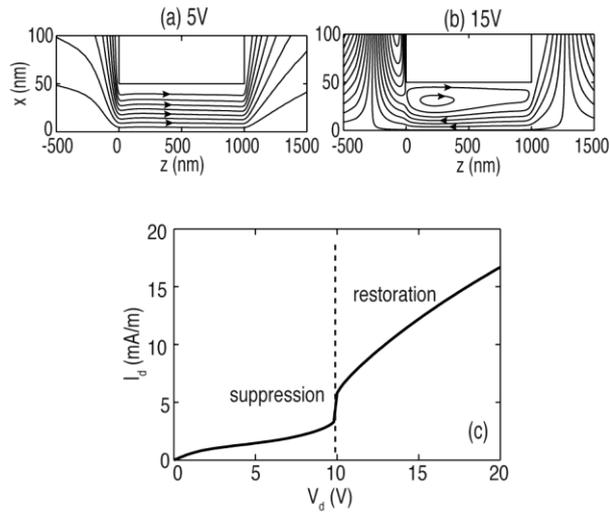

*Fig. 3: (a)&(b) Simulated fluid flow lines around the channel of the structure in Fig. 1a using the PNP-S model. Results are shown for moderate (5V) and high (15V) biases. Only the top half of the structure is shown considering the symmetry; (c) simulated current-voltage curve for this structure. The vertical line is a visual guide separating the conductance suppression and restoration stages.*

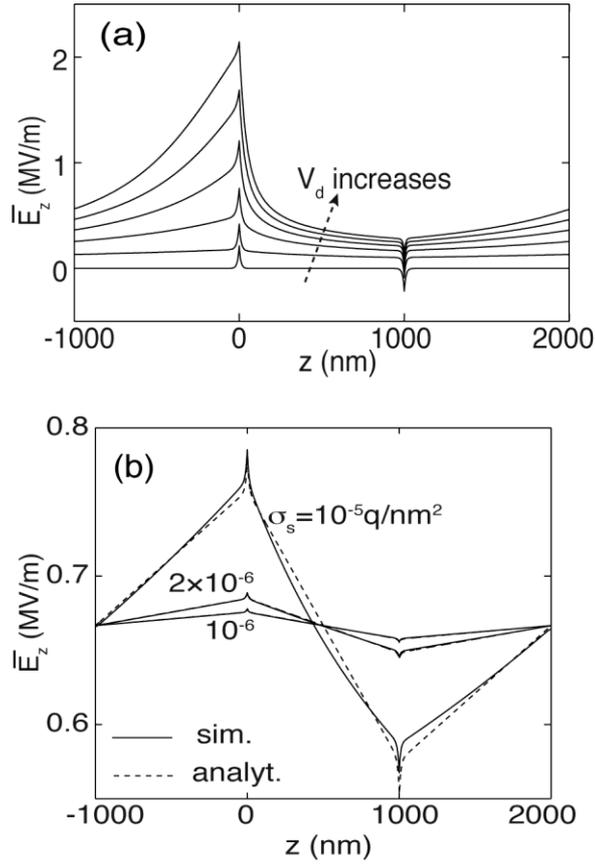

*Fig. 4: (a) Simulated distributions of transversally averaged longitudinal electric field, $\overline{E_z}$, for the simplified structure in Fig. 1b. The model is PNP. $\sigma_s$ is fixed at $10^{-2} q/nm^2$. $V_d$ increases from 0V to 2V at a step of 0.4V; (b) Distributions of the transversally averaged $\overline{E_z}$ for various surface charge densities. The model is PNP. Both simulation and analytical (Eq. 5) results are shown. $V_d$ is fixed at 2V.*

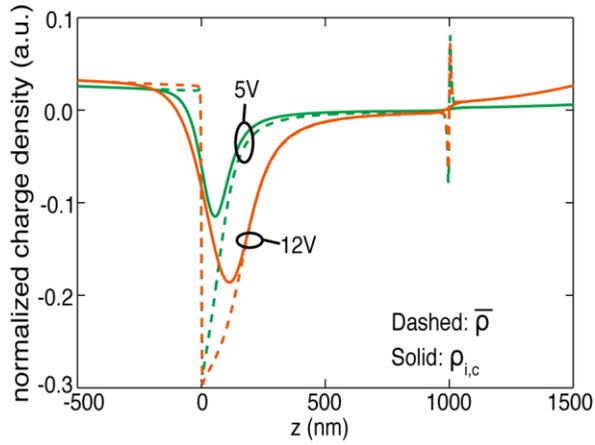

Fig. 5: Simulated distributions of normalized net charges for the simplified structure in Fig. 1b. The PNP model is used. Two quantities are plotted, the average net charges (including both ion and surface charges), $\bar{\rho}$, and the local ion charges at the centerline, $\rho_{i,c}$. Both quantities are normalized to $qc_0$. Results are shown for two biases, 5V and 12V.